\documentclass[12pt]{article}
\usepackage{amssymb}
\usepackage{graphicx}
\usepackage{graphics}
\usepackage{epsfig}
\usepackage{amsmath}
\usepackage{url}
\usepackage{multicol}

\newcommand{\be}{\begin{equation}}
\newcommand{\ee}{\end{equation}}
\newcommand{\ba}{\begin{align}}
\newcommand{\ea}{\end{align}}
\newcommand{\bg}{\begin{gather}}
\newcommand{\eg}{\end{gather}}
\newcommand{\bseq}{\begin{subequations}}
\newcommand{\eseq}{\end{subequations}}

\renewcommand{\tanh}{\mathop{\rm th}\nolimits}

\begin{document}

\title{
    Analytic Q-ball solutions and their stability\\ in a piecewise
    parabolic potential
  }

\author{
I.\,E.\;Gulamov$^{a}$, E.\,Ya.\;Nugaev$^b$\thanks{{\bf e-mail}:
emin@ms2.inr.ac.ru}, M.\,N.\;Smolyakov$^c$\thanks{{\bf e-mail}:
smolyakov@theory.sinp.msu.ru}
\\
$^a${\small{\em Physics Department, Lomonosov Moscow State
University,
}}\\
{\small{\em 119991, Moscow, Russia}}\\
$^b${\small{\em
Institute for Nuclear Research of the Russian Academy
of Sciences,}}\\
{\small{\em 60th October Anniversary prospect 7a, Moscow 117312,
Russia
}}\\
$^c${\small{\em Skobeltsyn Institute of Nuclear Physics, Lomonosov
Moscow State University,
}}\\
{\small{\em 119991, Moscow, Russia}}}

\date{}
\maketitle

\begin{abstract}
Explicit solutions for extended objects of a Q-ball type were
found analytically in a model describing complex scalar field with
piecewise parabolic potential in (3+1)- and (1+1)-dimensional
space-times. Such a potential provides a variety of solutions
which were thoroughly examined. It was shown that, depending on
the values of the parameters of the model and according to the
known stability criteria, there exist stable and unstable
solutions. The classical stability of solutions in
(1+1)-dimensional space-time was examined in the linear
approximation and it was shown explicitly that the spectrum of
linear perturbations around some solutions contains exponentially
growing modes while it is not so for other solutions.
\end{abstract}

\section{Introduction}
Q-ball is a nontopological soliton in theories with a global
symmetry \cite{Coleman:1985ki}. A simple example is the model with
one complex scalar field $\phi$ with $U(1)$-invariant potential
$V(\phi^*\phi)$ which satisfies several simple conditions derived
in \cite{Coleman:1985ki}. The standard solution for a Q-ball has
the form
\begin{equation}
\phi(t,\vec{x})=f(r){\it e}^{{\it i}\omega t},
\label{solution-form}
\end{equation}
where $\omega$ is a constant and $f(r)$ is a monotonically
decreasing (in a simplest case) spherically symmetric function,
tending to zero at spatial infinity. Different applications in
cosmology (see, for example, \cite{Gorbunov:2011zz} for review)
encourage investigation of specific features of such classical
extended objects. However, nonlinearity of classical field
equations is a serious obstacle for obtaining analytical solutions
even in theories with simple polynomial potentials (see, for
example, \cite{Anderson:1970et}), the exceptions like the one
presented in \cite{Rosen} (the scalar field potential of this
model makes it possible to examine analytically even the linear
perturbations above the Q-ball solution, see \cite{MarcVent}) are
very rare. Some results of qualitative and numerical analysis of
Q-ball properties can be found in \cite{Tsumagari:2008bv}.

To obtain an analytic solution the model with parabolic piecewise
potential
\begin{equation}
V(\phi^*\phi)=M^2\left[\phi^*\phi+2\epsilon v
(v-\sqrt{\phi^*\phi})\theta\left(\frac{\sqrt{\phi^*\phi}}{v}-1\right)\right]
\label{Theodorakis}
\end{equation}
was considered in \cite{Theodorakis:2000bz}. Here $M^2>0$,
$\theta$ is the Heaviside step function, $v$ is a parameter of the
model, $\epsilon$ is a positive dimensionless constant. Indeed, in
both regions $f>v$ and $f<v$ the corresponding equation of motion
is analytically solvable and the solutions are regular and smooth.
Matching conditions determine the point $r=R$ such that $f(R)=v$.
Inside the sphere with radius $R$ the value of the field $f$ is
larger than $v$ and outside this sphere $f<v$ and exponentially
tends to zero. Thus, the charge and the energy are localized
inside the ball with the center at the origin $r=0$.

\begin{figure}[!t]
\includegraphics[width=6.5in]{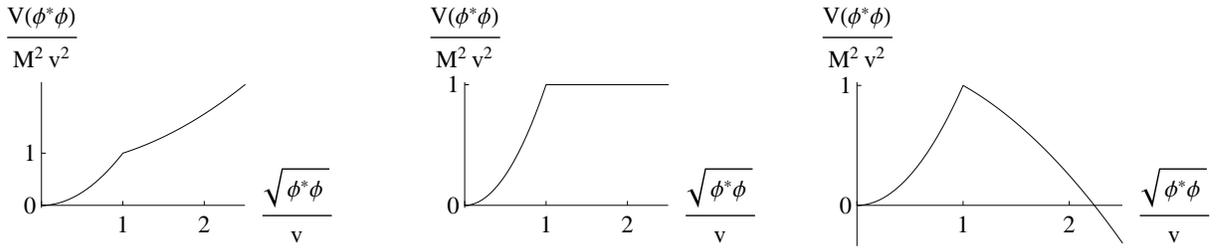}
\caption{The forms of the scalar field potential described by
Eq.~(\ref{potential}): $m^2>0$, $M/m=2$ (left plot); $m=0$ (middle
plot); $m^2<0$, $M/|m|=2$ (right plot).} \label{V}
\end{figure}

In a solvable model, one can find the whole spectrum of
excitations above a Q-ball solution, examine the modes responsible
for possible instabilities and, in addition, demonstrate
explicitly the validity of the stability conditions for Q-balls,
which can be found in \cite{Friedberg:1976me,LeePang,GSS}. Indeed,
one can examine linear perturbations above a Q-ball solution in
analogy with \cite{Anderson:1970et}, where instability of the
solution in a theory with potential of the form
$V=\kappa^2\phi^*\phi-\frac{\mu^2}{2}(\phi^*\phi)^2$ was
established numerically. The square root in potential
(\ref{Theodorakis}) is an obstacle for analytical consideration of
linear perturbations above the Q-ball solution. In this paper we
propose a model describing complex scalar field with potential of
the form
\begin{equation}
V(\phi^*\phi)=M^2\phi^*\phi\,\theta\left(1-\frac{\phi^*\phi}{v^2}\right)+(m^2\phi^*\phi+\Lambda)\theta\left(\frac{\phi^*\phi}{v^2}-1\right),
\label{potential}
\end{equation}
where $M^2>0$, $M^2>m^2$, $\theta$ is the Heaviside step function
with the convention $\theta(0)=\frac{1}{2}$, in theories with
three and one spatial dimensions. The constant $\Lambda$ provides
continuity of the potential at the point $\phi^*\phi=v^2$,
$\Lambda=v^2(M^2-m^2)$. This potential is presented in
Fig.~\ref{V}.
Of course, this piecewise potential should be considered as a
limiting case of some smooth and bounded potential.

Below we will examine Q-ball solutions in a model with potential
(\ref{potential}) in $(3+1)$- and $(1+1)$-dimensional space-times
and show that the model contains solutions possessing different
properties. As will be shown below, some solutions are stable,
some of them are not. According to \cite{Anderson:1970et},
examination of classical stability in the linear approximation
will be made for the $(1+1)$-dimensional case and stability
conditions, presented in \cite{Friedberg:1976me,LeePang,GSS}, will
be illustrated explicitly.

\section{Q-ball in four-dimensional space-time}
\subsection{Analytical Q-ball solution}
First let us consider (3+1)-dimensional space-time. The action of
the model has the form
\begin{equation}\label{act3D}
S=\int
d^4x\left[\partial_\mu\phi^*\partial^\mu\phi-M^2\phi^*\phi\,\theta\left(1-\frac{\phi^*\phi}{v^2}\right)-(m^2\phi^*\phi+\Lambda)\theta\left(\frac{\phi^*\phi}{v^2}-1\right)\right]
\end{equation}
with $\Lambda=v^2(M^2-m^2)$. We consider the standard form of
solution (\ref{solution-form}). The simplest regular spherically
symmetric solution to the equation of motion, coming from
(\ref{act3D}), takes the form
\begin{eqnarray}
f&=&B\frac{e^{-\sqrt{M^{2}-\omega^2}r}}{r},\qquad\textrm{for}\qquad
f^2<v^2,\\
f&=&A\frac{\sin({\sqrt{\omega^2-m^{2}}r})}{r},\qquad\textrm{for}\qquad
f^2>v^2.
\end{eqnarray}
It is clear that if $m^2>0$, then $m<|\omega|<M$; if $m^2=0$, then
$0<|\omega|<M$; otherwise $0\le|\omega|<M$. The continuity of the
solution and its first derivative at the point $r=R$ such that
$f(R)=v$ defines the coefficients $A$ and $B$, which read as
\begin{eqnarray}
B=\frac{vR}{e^{-\sqrt{M^{2}-\omega^2}R}},\\
A=\frac{vR}{\sin({\sqrt{\omega^2-m^{2}}R})}
\end{eqnarray}
and defines the value of the matching radius $r=R$:
\begin{equation}\label{R3D}
\tan\left(\sqrt{\omega^2-m^{2}}R\right)=-\frac{\sqrt{\omega^2-m^{2}}}{\sqrt{M^2-\omega^2}}.
\end{equation}
Since the parameter $R$ is positive, it should be taken to be
\begin{equation}\label{R3D-definition}
R=\frac{1}{\sqrt{\omega^2-m^{2}}}\left(\arctan\left(-\frac{\sqrt{\omega^2-m^{2}}}{\sqrt{M^2-\omega^2}}\right)+\pi\right).
\end{equation}
All the features of our solution can be expressed through this
parameter $R$ or, equivalently, through the frequency $\omega$.
Note that the presented solution is the simplest one -- the
absolute value of the function $f$ is a monotonically decreasing
function (which means that the charge density is a monotonic
function of $r$ as well), which is equal to $v$ only at the one
point $r=R$. Obviously, there may exist solutions which cross the
line $\sqrt{\phi^{*}\phi}=v$ several times and have nodes (i.e.
such that there exist points $r_{i}\ne\infty$: $f(r_{i})=0$).
Because of the simplicity of potential (\ref{potential}), such
solutions can also be found analytically. It would be interesting
to examine their properties in comparison with those of the
simplest solution presented above, but this topic lies beyond the
scope of the present paper.

One sees that the scalar field potential is unbounded from below
if $m^2<0$. Meanwhile, one can always consider a potential which
coincides with the one defined by (\ref{potential}) for
$\phi^*\phi\le{\tilde v}^2$ but changes its behavior and becomes
growing for $\phi^*\phi>{\tilde v}^2$, where $\tilde v>v$. In such
a case one should take only those Q-ball solutions for which the
maximum absolute value of the scalar field $|f(0)|\le \tilde v$.

Now let us examine the properties of the solution at hand. It is
not difficult to calculate the total charge and the total energy
of the Q-ball. They look as follows:
\begin{equation}\label{charge3D}
Q=-i\int d^3x\left(\phi^{*}\dot\phi-\dot\phi^{*}\phi\right)=4\pi
R^2\omega
v^2\left(\frac{(M^2-m^2)\left(R\sqrt{M^2-\omega^2}+1\right)}{\left(\omega^2-m^2\right)\sqrt{M^2-\omega^2}}\right),
\end{equation}
\begin{eqnarray}\label{energy3D}
E&=&\int
d^3x\left(\dot\phi^{*}\dot\phi+\partial_{i}\phi^{*}\partial_{i}\phi+V(\phi^{*}\phi)\right)=\\
\nonumber &=&4\pi
R^2\omega^2v^2\left(\frac{(M^2-m^2)\left(R\sqrt{M^2-\omega^2}+1\right)}{\left(\omega^2-m^2\right)\sqrt{M^2-\omega^2}}\right)+\frac{4\pi}{3}R^{3}v^{2}(M^{2}-m^{2}),
\end{eqnarray}
where $R$ is defined by (\ref{R3D-definition}). One can see that
$E=\omega Q+4\pi R^{3}v^{2}(M^{2}-m^{2})/3>\omega Q$. Note that
the inequality $E>\omega Q$ should hold for any Q-ball solution
(see simple derivation of this fact in Appendix~A). Another
interesting observation is that
\begin{equation}\label{QdefR}
Q=-v^{2}(M^{2}-m^{2})\frac{d\left(\frac{4\pi}{3}R^{3}\right)}{d\omega}.
\end{equation}
This observation allows one to make another cross-check of our
results. Indeed, let us differentiate equation (\ref{energy3D})
with respect to the charge $Q$:
\begin{equation}
\frac{dE}{dQ}=\omega+\frac{d\omega}{dQ}Q+\frac{4\pi}{3}v^{2}(M^{2}-m^{2})\frac{d(R^{3})}{dQ}=
\omega+\frac{d\omega}{dQ}Q+\frac{4\pi}{3}v^{2}(M^{2}-m^{2})\frac{d(R^{3})}{d\omega}\frac{d\omega}{dQ}.
\end{equation}
Using equation (\ref{QdefR}) we easily get the well-known relation
\cite{Tsumagari:2008bv,Friedberg:1976me,LeePang}
\begin{equation}\label{dEdQomega}
\frac{dE}{dQ}=\omega.
\end{equation}
which should hold for any Q-ball solution.

\subsection{$m^2>0$}
\begin{figure}[!h]
\includegraphics[width=6.5in]{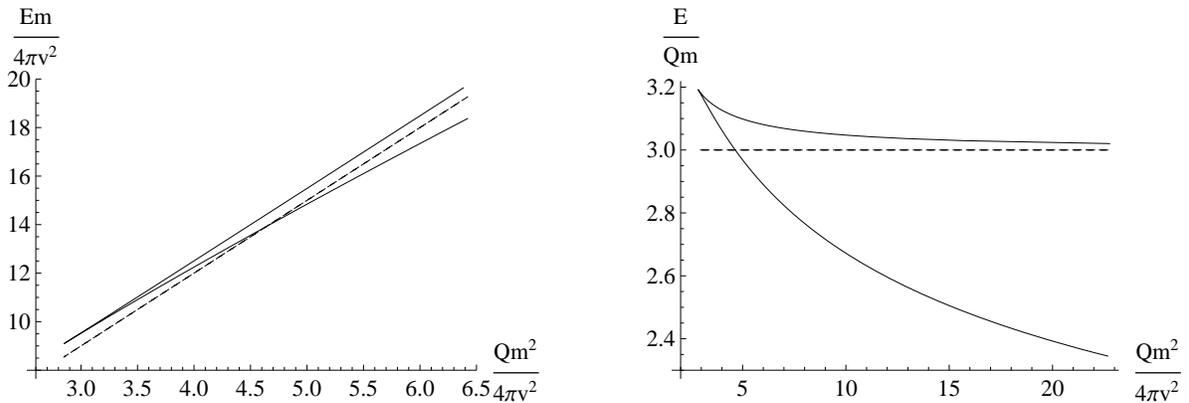}
\caption{$E(Q)$ (left plot) and $E(Q)/Q$ (right plot) for the
Q-ball in (3+1)-dimensional theory, $m^2>0$, $M/m=3$. The dashed
line corresponds to free particles with $E=MQ$.} \label{3D}
\end{figure}

First, we consider the case $m^2>0$. The $E(Q)$ and $E(Q)/Q$
diagrams for this case for $M/m=3$ (here and below, without loss
of generality, we consider $\omega>0$) are presented in
Fig.~\ref{3D}. We see that there are two different branches in
this figure, which correspond to solutions which are localized in
different ways. Localization of solutions with larger energy is
due to the exponential suppression outside the core $[0,R]$ and we
will refer to it as a "wide" branch. The size of solutions with
smaller energy is just $R$ (see below) and most of their charge is
concentrated inside the core $[0,R]$. We will refer to these
solutions as a "narrow" branch.
We see that there is a solution with nonzero minimal charge and
energy. This result is very similar to those obtained in
\cite{Friedberg:1976me,Alford:1987vs} where completely different
scalar field potentials were utilized. Indeed, one can see that
the left plot presented in Fig.~\ref{3D} looks the same as the
corresponding plots in Fig.~2 and Fig.~3 of
\cite{Friedberg:1976me}. The right plot in Fig.~\ref{3D} has the
same form as the one presented in \cite{Alford:1987vs}. Of course,
these results are also very similar to the results obtained in
\cite{Theodorakis:2000bz}.

The quantitative difference between branches can be expressed
through the parameter
\begin{equation}
g=\sqrt{\frac{M^2-\omega^2}{\omega^2-m^2}} \label{g1dim}.
\end{equation}
In the case of large $Q$ two different branches, presented in
Fig.\ref{3D}, correspond to $g\gg 1$ and $g\ll 1$. For large $Q$
and $M\sim m$ the properties of the solutions are summarized in
Table~1.

\begin{center}
\begin{tabular}{|c|c|c|}
\hline
Type of solution for large $Q$& Wide branch & Narrow branch\\
\hline
Asymptotics  & $g\ll 1$ & $g \gg 1$\\
\hline
Soliton radius $L$ & $\sim\frac{1}{g\sqrt{M^2-m^2}}$ & $ \sim R\sim \frac{\pi g}{\sqrt{M^2-m^2}}$\\
\hline
Soliton energy $E$ & $\sim v^2L$ & $\sim v^2M^4L^5$\\
\hline
Soliton charge $Q$ & $\sim v^2L/M$ & $\sim v^2M^3L^5$\\
\hline
\end{tabular}

\vspace{0.5cm} Table~1. Properties of the Q-ball solutions for
large Q; $m^{2}>0$; (3+1)-dimensional space-time.
\end{center}

One can express all values in terms of large $Q$. For example, if
$m\sim M$ the size of wide Q-ball is $L_{w}\sim \frac{QM}{v^2}$
and the size of narrow Q-ball is $L_{n}\sim
(\frac{Q}{M^{3}v^2})^{1/5}$. This difference is the origin for the
titles of the branches -- indeed, for large $Q$ and for $M\sim v$
we have $L_{w}\gg L_{n}$. Solutions of the wide branch are very
similar to the Q-clouds of \cite{Alford:1987vs}.

It should be mentioned that the usual thin-wall approximation can
not be used for solutions presented above. Indeed, for the stable
branch even in the limit $\omega\to m$ (in this case the solution
most rapidly falls off in the region $r>R$) the solution
considerably differs from a constant in the region $[0,R]$.

\subsection{$m=0$}
Now we turn to the second case $m=0$. The scalar field potential
in this case contains flat directions, such type of potentials
arise in supersymmetric theories.
\begin{figure}[h!]
\includegraphics[width=6.5in]{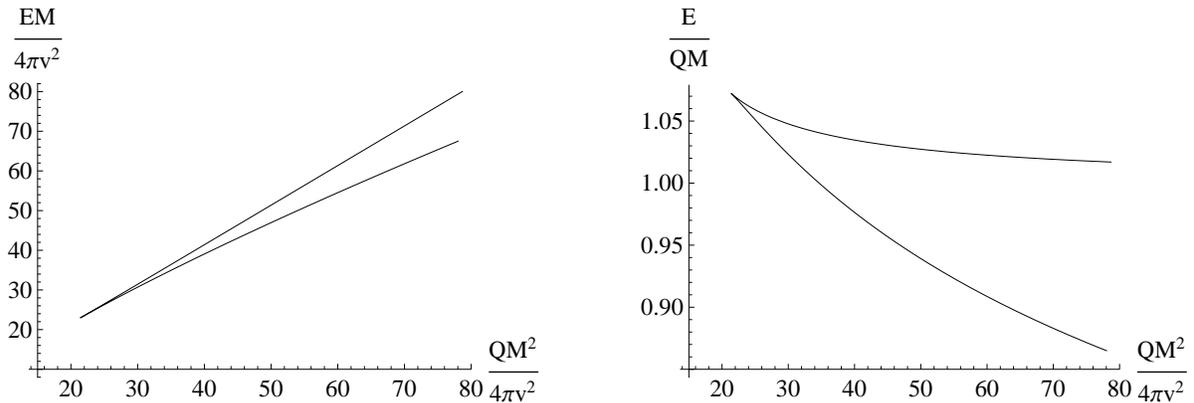}
\caption{$E(Q)$ (left plot) and $E(Q)/Q$ (right plot) for the
Q-ball in (3+1)-dimensional theory, $m=0$.} \label{3D0}
\end{figure}
As can be seen from Fig.~\ref{3D0}, the properties of the Q-ball
solutions in this case look very similar to those discussed in the
previous subsection. Meanwhile, there is a considerable difference
between the cases $m^2>0$ and $m=0$. In the first case solutions
with large charge $Q$ on the lower branch have the following
energy-charge dependence: $E\sim Q$. In the second case $m=0$ one
obtains from (\ref{charge3D}) and (\ref{energy3D}) for large $Q$
(i.e for $\omega\to 0$) the following energy-charge dependence:
$E\sim Q^{\frac{3}{4}}$. The latter relation coincides with the
general estimate for potentials of this type, which can be found
in \cite{Gorbunov:2011zz}.

\subsection{$m^2<0$}
\begin{figure}[h!]
\includegraphics[width=6.5in]{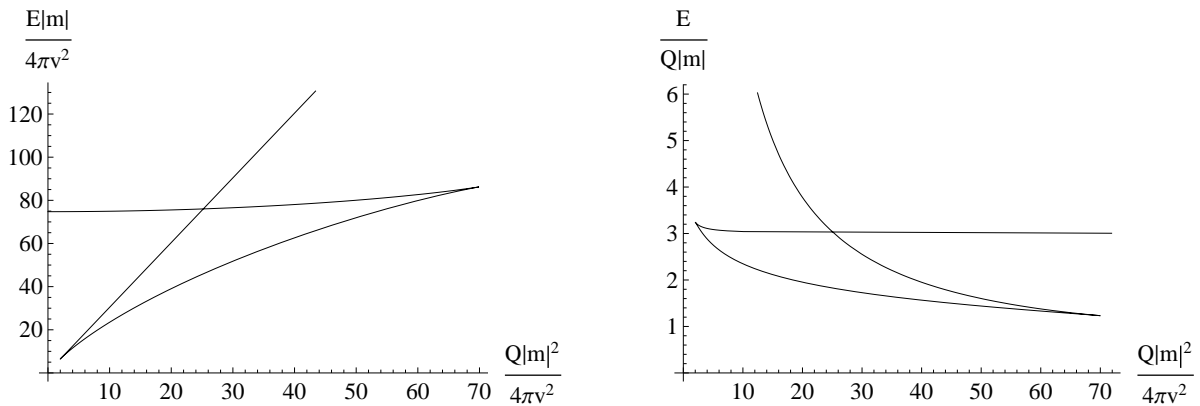}
\caption{$E(Q)$ (left plot) and $E(Q)/Q$ (right plot) for the
Q-ball in (3+1)-dimensional theory, $m^2<0$, $M/|m|=3$.}
\label{3D-less-0}
\end{figure}

\begin{figure}[h!]
\includegraphics[width=6.5in]{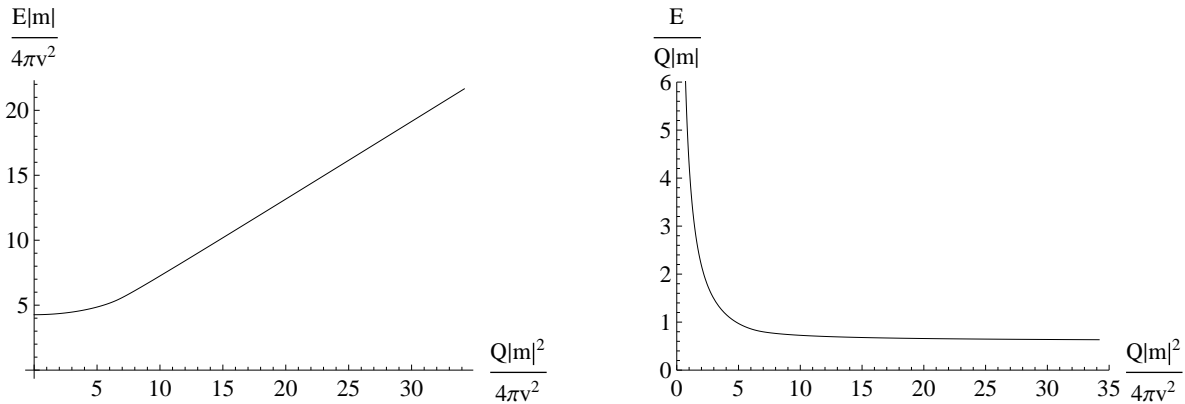}
\caption{$E(Q)$ (left plot) and $E(Q)/Q$ (right plot) for the
Q-ball in (3+1)-dimensional theory, $m^2<0$, $M/|m|=0.6$.}
\label{3D-less-0-2}
\end{figure}

The third case $m^2<0$ appears to be completely different from the
previous cases. First, there are two different phases -- solutions
in the first phase contain three branches in the $E(Q)$ diagram
(see Fig.~\ref{3D-less-0}),
whereas another phase contains only one branch (see
Fig.~\ref{3D-less-0-2}).
The transition between the phases occurs at $\frac{M}{|m|}\approx
0.728$.

We see that in Fig.~\ref{3D-less-0} there exist three branches,
two of them intersect each other. There also exists nontrivial
(i.e. having nonzero energy) solution with the zero charge $Q$. An
analogous case was described in \cite{Theodorakis:2000bz} for the
appropriate range of the parameters of the scalar field potential.
Note that though the scalar field potential in
\cite{Theodorakis:2000bz} is bounded from below for any values of
the parameters, whereas our potential is unbounded from below for
$m^2<0$, similar Q-ball solutions having three branches arise in
both cases. An interesting observation is that the part of $E(Q)$
dependence containing two branches, i.e. the part starting from
$\omega=0$ and ending at $\omega=\omega_{c}<M$, where $\omega_{c}$
corresponds to the solution with minimal (this minimum is local in
general) energy, resembles the $E(Q)$ dependence found in
\cite{MarcVent} in the model with a completely different
potential. The difference between our case and the case discussed
in \cite{MarcVent} is that the $E(Q)$ dependence of
\cite{MarcVent} has $\omega_{c}\to\infty$ and the corresponding
branch ends at the point $(0;0)$ on the $Q,E$ plane.

The $E(Q)$ dependence of another phase (presented in
Fig.~\ref{3D-less-0-2}) is very similar to the one of the model
presented in \cite{Anderson:1970et}.

With the help of the explicit solution it can be shown that the
thin-wall approximation also does not describe the case $m^2<0$.

\subsection{Stability of the solutions}
Now let us discuss the stability of the Q-ball solutions presented
above. The first type of stability is the quantum mechanical
stability. We will focus on the case $m^2>0$. One sees from
Fig.~\ref{3D} that the lower branch crosses the line $E=MQ$,
corresponding to free particles with the rest mass $M$, at some
charge, say $Q_{x}$. Thus, for the region of charges $Q>Q_{x}$ the
energy of the Q-ball is smaller than the energy of free particles
and thus such a Q-ball is quantum mechanically stable. In the
region $Q_{min}<Q<Q_{x}$, where $Q=Q_{min}\approx 2.85\frac{4\pi
v^{2}}{m^{2}}$ for $M/m=3$, the Q-ball solution is unstable with
respect to radiation of free particles. Q-balls from the upper
"wide" branch are always unstable from this point of view.

Now let us consider the Q-ball fission. It is known that Q-balls
are stable against fission if $d^{2}E/dQ^{2}<0$ (although this
condition is known, we present a simple justification of this fact
in Appendix~B). This relation holds for the lowest branches in all
three cases (for the case $m^2<0$ this is valid for the phase with
$\frac{M}{|m|}\gtrsim 0.728$). Indeed,
$\frac{d^{2}E}{dQ^{2}}=\frac{d\omega}{dQ}$ (see Eq.
(\ref{dEdQomega})), whereas for the lowest branches
$\frac{dQ}{d\omega}<0$ because the charge increases while $\omega$
decreases and thus $d^{2}E/dQ^{2}<0$. This can also be seen from
Fig.~\ref{3D}, Fig.~\ref{3D0} and Fig.~\ref{3D-less-0}.

The last type of stability, which can be discussed here, is the
classical stability, i.e. stability with respect to small
perturbations. The classical stability criterion proposed in
\cite{Friedberg:1976me,LeePang} implies that a Q-ball solution is
classically stable if $\frac{dQ}{d\omega}<0$ (for $\omega>0$ and
$Q>0$, which is exactly our case), a mathematically rigorous proof
of this fact can be found in \cite{GSS} (see also references
therein). This relation, as it was shown above, holds for the
lowest branches in all three cases (for the case $m^2<0$ this is
valid for the phase with $\frac{M}{|m|}\gtrsim 0.728$) and thus we
can conclude that Q-balls from these branches are classically
stable. An interesting observation for the case $m^2<0$,
$\frac{M}{|m|}\gtrsim 0.728$ is that classically stable Q-ball
solutions lie between solutions with locally minimal and locally
maximal charges, i.e. there exist stable solution with minimal
charge (and energy) and stable solution with maximal charge (and
energy).

Meanwhile, it would be interesting to consider linear
perturbations explicitly in order to examine the upper branches
and their possible instability modes. A complete analytical
analysis of linearized theory in the (3+1)-dimensional case seems
to be a rather complicated task and it lies beyond the scope of
this paper. Nevertheless, we performed analysis of perturbations
in the linear approximation in a simpler model in
(1+1)-dimensional space-time for the cases $m^2>0$ and $m^2<0$.
The physical properties of the (1+1)-dimensional model are very
similar to those in the (3+1)-dimensional case and we think that
results obtained in such a simplified case can be applied to the
(3+1)-dimensional case as well.

\section{Q-ball in two-dimensional space-time}
\subsection{Analytical solution and its properties}
Now the action takes the form
\begin{equation}\label{act1D}
S=\int
dtdz\left(\partial_\mu\phi^*\partial^\mu\phi-M^2\phi^*\phi\,\theta\left(1-\frac{\phi^*\phi}{v^2}\right)-(m^2\phi^*\phi+\Lambda)\theta\left(\frac{\phi^*\phi}{v^2}-1\right)\right)
\end{equation}
with $\Lambda=v^2(M^2-m^2)$. Again we consider the standard ansatz
\[
\phi(t,z)=f(z){\rm e}^{i\omega t}
\]
with a dimensionless even function $f(z)=f(-z)$. At the points
$z=\pm R$ function $f(z)$ is equal to $v$ and there are
discontinuities in the ordinary differential equation on $f(z)$:
\begin{equation}
-\partial^2_z
f-\omega^2f+M^2\theta\left(1-\frac{f^2}{v^2}\right)f+m^2\theta\left(\frac{f^2}{v^2}-1\right)f=0.
\label{eq1dim}
\end{equation}
Inside the interval $(-R,R)$ the function $|f|$ is larger than $v$
and outside this interval it falls off exponentially to zero. The
corresponding solution to equation of motion (\ref{eq1dim}) takes
the form
\begin{eqnarray}
f&=&ve^{-\sqrt{M^{2}-\omega^2}(|z|-R)},\qquad\textrm{for}\qquad
f^2<v^2,\quad |z|\ge R\\
f&=&v\frac{\cos({\sqrt{\omega^2-m^{2}}z})}{\cos({\sqrt{\omega^2-m^{2}}R})},\qquad\textrm{for}\qquad
f^2>v^2,\quad |z|<R
\end{eqnarray}
with $R$ defined as
\begin{equation}\label{R1D}
R=\frac{1}{\sqrt{\omega^2-m^2}}\arctan\left(\frac{\sqrt{M^2-\omega^2}}{\sqrt{\omega^2-m^2}}\right).
\end{equation}
The total charge and the total energy look as follows:
\begin{equation}\label{charge1D}
Q=2v^2\omega
\left(\frac{(M^2-m^2)\left(R\sqrt{M^2-\omega^2}+1\right)}{\left(\omega^2-m^2\right)\sqrt{M^2-\omega^2}}\right),
\end{equation}
\begin{equation}\label{energy1D}
E=2v^2
\omega^2\left(\frac{(M^2-m^2)\left(R\sqrt{M^2-\omega^2}+1\right)}{\left(\omega^2-m^2\right)\sqrt{M^2-\omega^2}}\right)+2v^2(M^{2}-m^{2})R,
\end{equation}
where $R$ is defined by (\ref{R1D}). Again one sees that $E=\omega
Q+2v^2(M^{2}-m^{2})R>\omega Q$. It is straightforward to show
that, in analogy with the (3+1)-dimensional case, the following
relation holds:
\begin{equation}
Q=-v^{2}(M^{2}-m^{2})\frac{d(2R)}{d\omega}.
\end{equation}
Using this relation we easily obtain (\ref{dEdQomega}).

\begin{figure}[h!]
\includegraphics[width=6.5in]{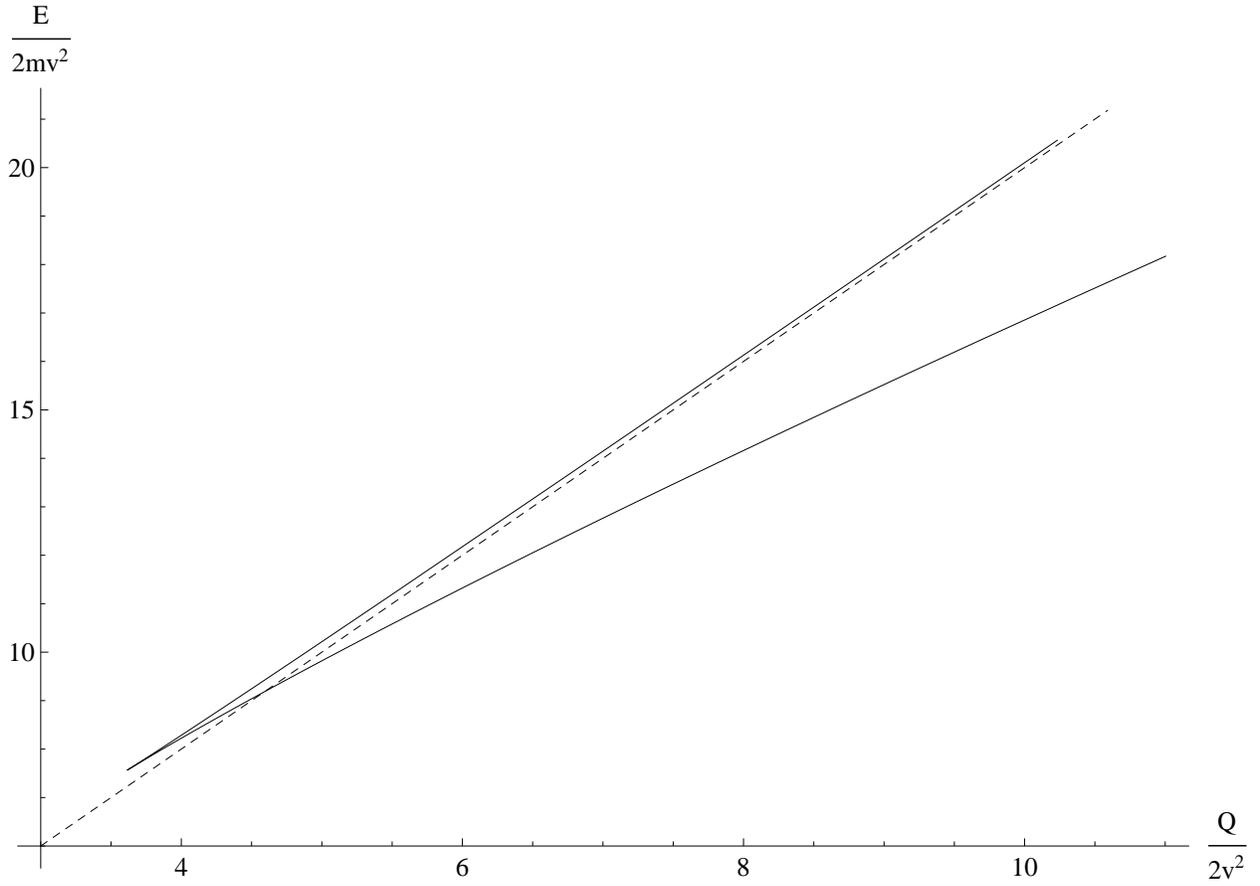}
\caption{$E(Q)$ for the Q-ball in (1+1)-dimensional theory (solid
line) and for free particles with $E=MQ$ (dashed line), $m^2>0$,
$M/m=2$. } \label{E_Q1dim}
\end{figure}

First we consider the simpler case $m^2>0$. The function $E(Q)$
for $M/m=2$ is presented in Fig.\,\ref{E_Q1dim}.
\begin{figure}[h!]
\includegraphics[width=6.5in]{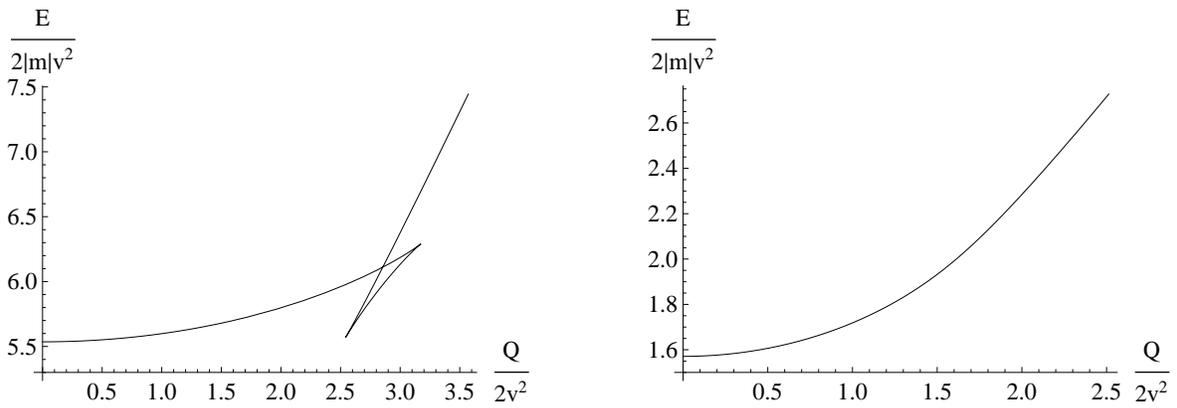}
\caption{$E(Q)$ for the Q-ball in (1+1)-dimensional theory,
$m^2<0$, $M/|m|=2$ (left plot) and $M/|m|=1$ (right plot). }
\label{E_Q1dim-2}
\end{figure}
As in the $(3+1)$-dimensional case, there exist two different
branches, their properties again can be expressed through the
parameter $g$ defined by (\ref{g1dim}) and the effective size $L$.
The differences between solutions are summarized in Table~2.
\begin{center}
\begin{tabular}{|c|c|c|}
\hline
Type of solution for large $Q$& Wide branch & Narrow branch\\
\hline
Asymptotics  & $g\ll 1$ & $g \gg 1$\\
\hline
Soliton size $L$ & $\sim\frac{1}{g\sqrt{M^2-m^2}}$ & $ \sim R\sim \frac{\pi g}{2\sqrt{M^2-m^2}}$\\
\hline
Soliton energy $E$ & $\sim v^2M^2L$ & $\sim v^2M^4L^3$\\
\hline
Soliton charge $Q$ & $\sim v^2ML$ & $\sim v^2M^3L^3$\\
\hline
\end{tabular}

\vspace{0.5cm} Table~2. Properties of the Q-ball solutions for
large Q; $m^{2}>0$; (1+1)-dimensional space-time.
\end{center}

The case $m^2<0$ is also very similar to the one in the
$(3+1)$-dimensional case. Again there are two phases, presented in
Fig.~\ref{E_Q1dim-2}. The transition between the phases in
(1+1)-dimensional space-time occurs at $\frac{M}{|m|}\approx
1.496$.

\subsection{Perturbations and classical stability analysis}
Now let us consider the linearized theory above the Q-ball
solution presented in the previous subsection. We will be
interested mainly in exponentially growing modes which indicate
the existence of classical instability. We consider small
perturbations above the classical solution of the form
\[
\phi(t,z)={\rm e}^{i\omega t}f(z)+{\rm e}^{i\omega t}\eta(t,z).
\]
Formally, the linearized equation of motion takes the form
\begin{equation}
-\ddot\eta-2i\omega\dot\eta+\omega^2\eta+\eta''-\left(M^2+(m^2-M^2)\theta(f^2-v^2)\right)\eta-(m^2-M^2)\delta(f^2-v^2)
f^2\left(\eta+\eta^{*}\right)=0,
\end{equation}
which can be easily brought to the form
\begin{equation}\label{perturb}
-\ddot\eta-2i\omega\dot\eta+\omega^2\eta+\eta''-\left(M^2+(m^2-M^2)\theta(f^2-v^2)\right)\eta-F\delta(|z|-R)
\left(\eta+\eta^{*}\right)=0
\end{equation}
where
\begin{equation}
F=F(\omega)=\frac{m^2-M^2}{2\sqrt{M^2-\omega^2}}.
\end{equation}
It is worth mentioning that the mixing between $\eta$ and $\eta^*$
occurs for potential (\ref{potential}) only at the points $z=\pm
R$ through the term with the $\delta$ function and this fact
really simplifies the consideration.

There are two obvious solutions to equation (\ref{perturb}). The
first one is the translational mode
\begin{equation}
\eta\sim f'(z)
\end{equation}
and the second one corresponds to the existence of the $U(1)$
global symmetry
\begin{equation}
\eta\sim if(z).
\end{equation}

In order to find other possible solutions in the linearized theory
we consider the standard ansatz for perturbations (see, for
example, \cite{Anderson:1970et,MarcVent})
\begin{equation}\label{subst-lin}
\eta=\psi_{1}(z){e}^{i\gamma t}+\psi_{2}^*(z){e}^{-i\gamma^* t}.
\end{equation}
Substituting this decomposition into (\ref{perturb}) one obtains
the equations
\begin{equation}
\left\{
 \begin{array}{rcl}
  \left[-\partial^2_z+U(z)+F\delta(|z|-R)\right]\psi_{1}+F\delta(|z|-R)\psi_{2}&=&(\omega+\gamma)^2\psi_{1},\\
  \left[-\partial^2_z+U(z)+F\delta(|z|-R)\right]\psi_{2}+F\delta(|z|-R)\psi_{1}&=&(\omega-\gamma)^2\psi_{2},\\
 \end{array}
\right. \label{stab_system}
\end{equation}
where $U(z)=M^2\theta(|z|-R)+m^2\theta(R-|z|)$. It should be noted
that formally the case $\textrm{Re}\gamma=0$ should be considered
separately, because in this case there is no separation of the
equation (\ref{perturb}) into terms proportional to
$e^{i(\textrm{Re}\gamma) t}$ and $e^{-i(\textrm{Re}\gamma) t}$,
which results in two different equations presented above. But it
appears that the equation for the spectrum, which will be obtained
from (\ref{stab_system}), is also valid for the case
$\textrm{Re}\gamma=0$.

First, let us consider normalized solutions to equations
(\ref{stab_system}). They take the form
\begin{equation}
\eta(z)=e^{i\gamma t}e^{i\mu_{1}z}a_{1}+e^{-i\gamma^{*}
t}e^{i\tilde\mu_{1}z}b_{1},\qquad \textrm{Im}\mu_{1}> 0,\quad
\textrm{Im}\tilde\mu_{1}> 0
\end{equation}
with $(\gamma+\omega)^{2}=M^{2}+\mu_{1}^{2}$ and
$(-\gamma^{*}+\omega)^{2}=M^{2}+\tilde\mu_{1}^{2}$ for $|z|>R$ and
\begin{equation}
\eta(z)=e^{i\gamma t}\left(e^{i\mu_{2}z}a_{2}+e^{-i\mu_{2}z}\tilde
a_{2}\right)+e^{-i\gamma^{*}t}\left(e^{i\tilde\mu_{2}z}b_{2}+e^{-i\tilde\mu_{2}z}\tilde
b_{2}\right)
\end{equation}
with $(\gamma+\omega)^{2}=m^{2}+\mu_{2}^{2}$ and $
(-\gamma^{*}+\omega)^{2}=m^{2}+\tilde\mu_{2}^{2}$ for $|z|\le R$.
Note that if the conditions $\textrm{Im}\mu_{1}>0$,
$\textrm{Im}\tilde\mu_{1}>0$ are not fulfilled, the solutions do
not fall off at spatial infinity and thus such solutions are not
normalizable.

It is convenient to consider modes which are even and odd in $z$
separately from the very beginning. In the first case we have
$\eta'|_{z=0}=0$. The matching condition at the point $z=R$,
coming from (\ref{perturb}), looks as follows:
\begin{equation}\label{matching}
\eta'|_{z=R+0}-\eta'|_{z=R-0}=F(\eta+\eta^{*})|_{z=R}.
\end{equation}
The condition (\ref{matching}) together with the continuity of
$\eta$ at the point $z=R$ generate the characteristic equation on
$\gamma$. After some calculations (which are straightforward but
quite tedious and we do not present the details of calculations
here), we can get the following characteristic equation for the
spectrum:
\begin{equation}\label{spectrum}
iF\Bigl(G(\mu_{1},\mu_{2})Y(-\tilde\mu_{2}^{*})+G(-\tilde\mu_{1}^{*},-\tilde\mu_{2}^{*})Y(\mu_{2})\Bigr)-
G(\mu_{1},\mu_{2})G(-\tilde\mu_{1}^{*},-\tilde\mu_{2}^{*})=0,
\end{equation}
where
$$
G(\mu_{1},\mu_{2})=\mu_{2}\left(1-e^{-i2\mu_{2}R}\right)-\mu_{1}\left(1+e^{-i2\mu_{2}R}\right),
$$
$$
Y(\mu_{2})=1+e^{-i2\mu_{2}R}.
$$
Note that equation (\ref{spectrum}) can be used only if the
conditions
\begin{eqnarray}\label{cond1}
&&\textrm{Im}\gamma\ne 0,\\
\label{cond2}&&\textrm{Im}\gamma=0\quad \textrm{and} \quad
\omega-M<\textrm{Re}\gamma<M-\omega
\end{eqnarray}
are fulfilled. In this case $\eta$ falls off exponentially with
$z\to\pm\infty$ and we have normalized solutions. Otherwise (i.e.
if $\textrm{Im}\gamma= 0$ and $\textrm{Re}\gamma\ge M-\omega$ or
$\textrm{Re}\gamma\le -M+\omega$) one should consider a more
general form of perturbations in order to describe modes from the
continuous spectrum.

Analogous calculations can be made for the odd modes, for which
$\eta|_{z=0}=0$. The matching condition again has the form
(\ref{matching}) and we get equation (\ref{spectrum}), but now
with
$$
G(\mu_{1},\mu_{2})=\mu_{1}\left(1-e^{-i2\mu_{2}R}\right)-\mu_{2}\left(1+e^{-i2\mu_{2}R}\right),
$$
$$
Y(\mu_{2})=-1+e^{-i2\mu_{2}R}.
$$
These formulas also can be used only if (\ref{cond1}) or
(\ref{cond2}) are fulfilled.

It is interesting to note that, according to (\ref{cond1}) and
(\ref{cond2}), the unstable mode (if exists) is normalized. As
already mentioned, the non-normalized modes from the continuous
spectrum exist if $\textrm{Im}\gamma=0$ and if
$M-\omega\le\textrm{Re}\gamma$ or $\textrm{Re}\gamma\le\omega-M$.

Now let us turn to an examination of the discrete spectrum of our
model. First, it is necessary to note that there exist obvious
solutions to equation (\ref{spectrum}) in the case of the odd
modes. Indeed, if $\textrm{Im}\gamma=0$, these solutions to
(\ref{spectrum}) are simply $\mu_{2}=0$ or $\tilde\mu_{2}=0$,
which leads to $\textrm{Re}\gamma=-\omega-m$,
$\textrm{Re}\gamma=-\omega+m$, $\textrm{Re}\gamma=\omega+m$, and
$\textrm{Re}\gamma=\omega-m$. But it can be shown that for these
values of $\gamma$ the only solution to the initial system of
linearized equations of motion is $\eta\equiv 0$. Thus, these
roots are unphysical. Note that depending on the value of $M/m$
these roots corresponding to unphysical modes may formally lie in
the continuous spectrum.

\begin{figure}[pH]
\includegraphics[width=6.5in]{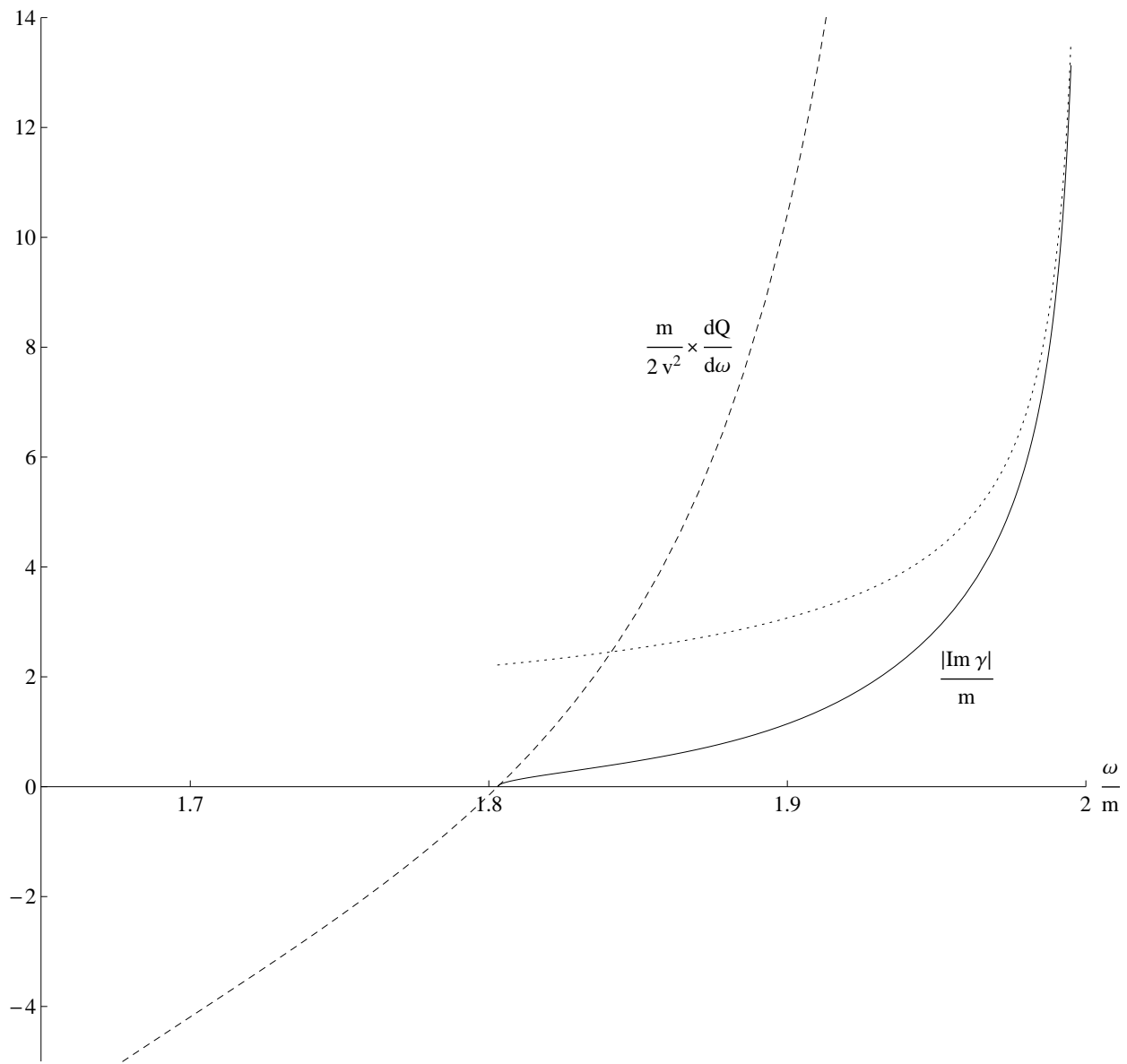}
\caption{The value of $|{\rm Im} \gamma|$ as a function of
$\omega$ for the case $m^2>0$, $M/m=2$. Solid line -- numerical
calculations; dotted line -- the function $\xi|F(\omega)|/m$. The
dashed line corresponds to the function $\frac{dQ}{d\omega}$.}
\label{gamma_Q}
\end{figure}

Equation (\ref{spectrum}) was examined numerically for $M/m=2$. We
have failed to find solutions to (\ref{spectrum}) for
$\textrm{Re}\gamma\ne 0$ and $\textrm{Im}\gamma\ne 0$. For the
case $\textrm{Im}\gamma=0$ we have found that solutions to
equation (\ref{spectrum}) may exist (or may not exist) depending
on the value of $\omega$. The case $\textrm{Re}\gamma=0$ and
$\textrm{Im}\gamma\ne 0$ was examined separately for different
values of $M/m$. We have failed to find any exponentially growing
odd mode. For $\omega>\omega_{c}$, where
$\omega_{c}=\omega_{c}\left(M/m\right)$ is the frequency of the
Q-ball solution corresponding to the minimal charge and energy
(for $M/m=2$ this frequency is $\omega_{c}\approx 1.803\, m$),
only one exponentially growing even mode was found (it is evident
that modes with nonzero imaginary part of $\gamma$ correspond to
classical instability). For $\omega<\omega_{c}$ we have failed to
find any exponentially growing even mode. Thus, the numerical
analysis shows that the upper ("wide") branch in
Fig.~\ref{E_Q1dim} is always classically unstable, whereas the
lower ("narrow") branch, for which $\frac{dQ}{d\omega}<0$, is
always classically stable, which coincides with the classical
stability criterion of \cite{Friedberg:1976me,LeePang}. An example
of a nontrivial numerical solution for ${\rm Im}\gamma$ for even
excitations together with the function $\frac{dQ}{d\omega}$ is
presented in Fig.~\ref{gamma_Q}.
\begin{figure}[h!]
\includegraphics[width=6.5in]{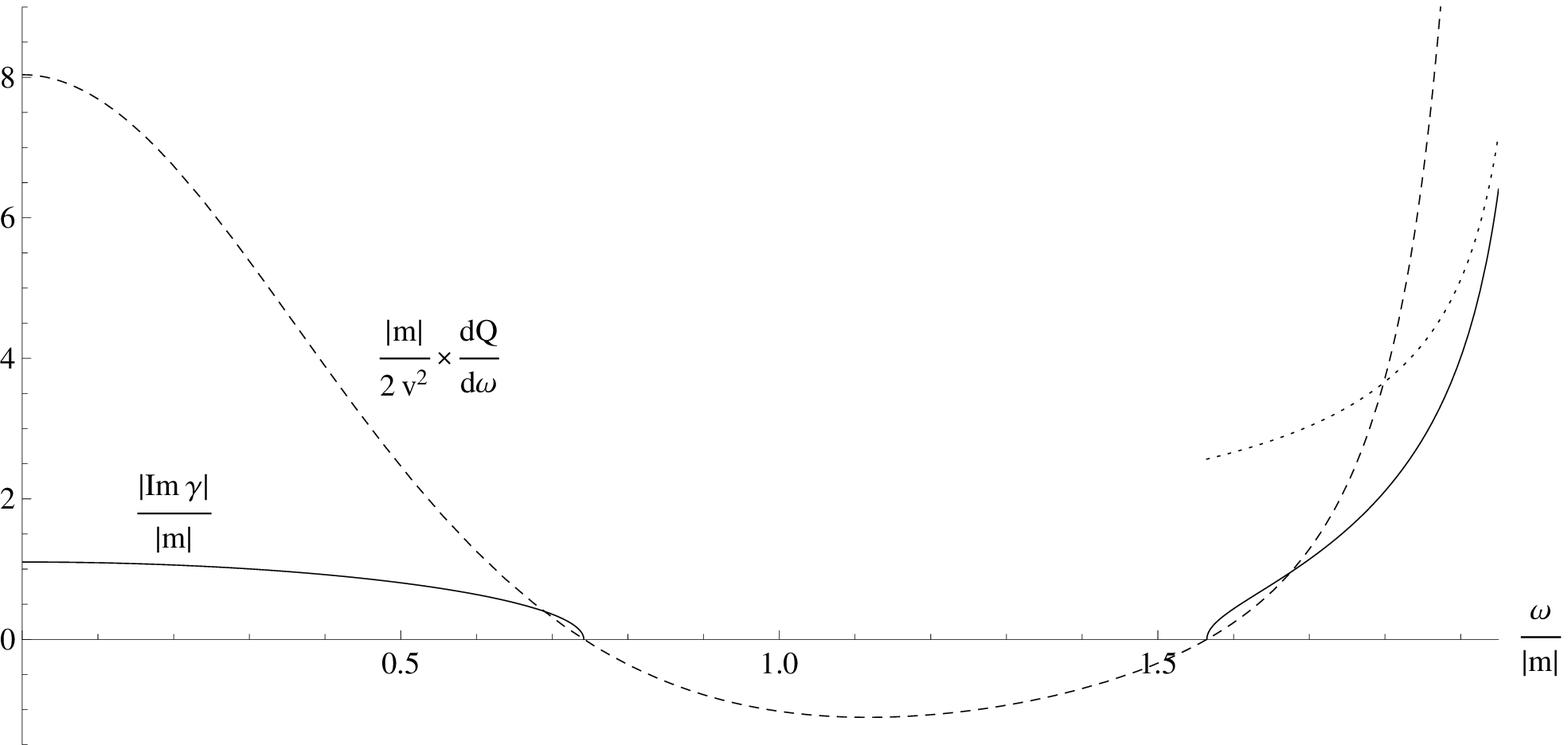}
\caption{The value of $|{\rm Im} \gamma|$ as a function of
$\omega$ for the case $m^2<0$, $M/|m|=2$. Solid line -- numerical
calculations; dotted line -- the function $\xi|F(\omega)|/|m|$.
The dashed line corresponds to the function $\frac{dQ}{d\omega}$.}
\label{gamma_Q1}
\end{figure}
\begin{figure}[h!]
\includegraphics[width=6.5in]{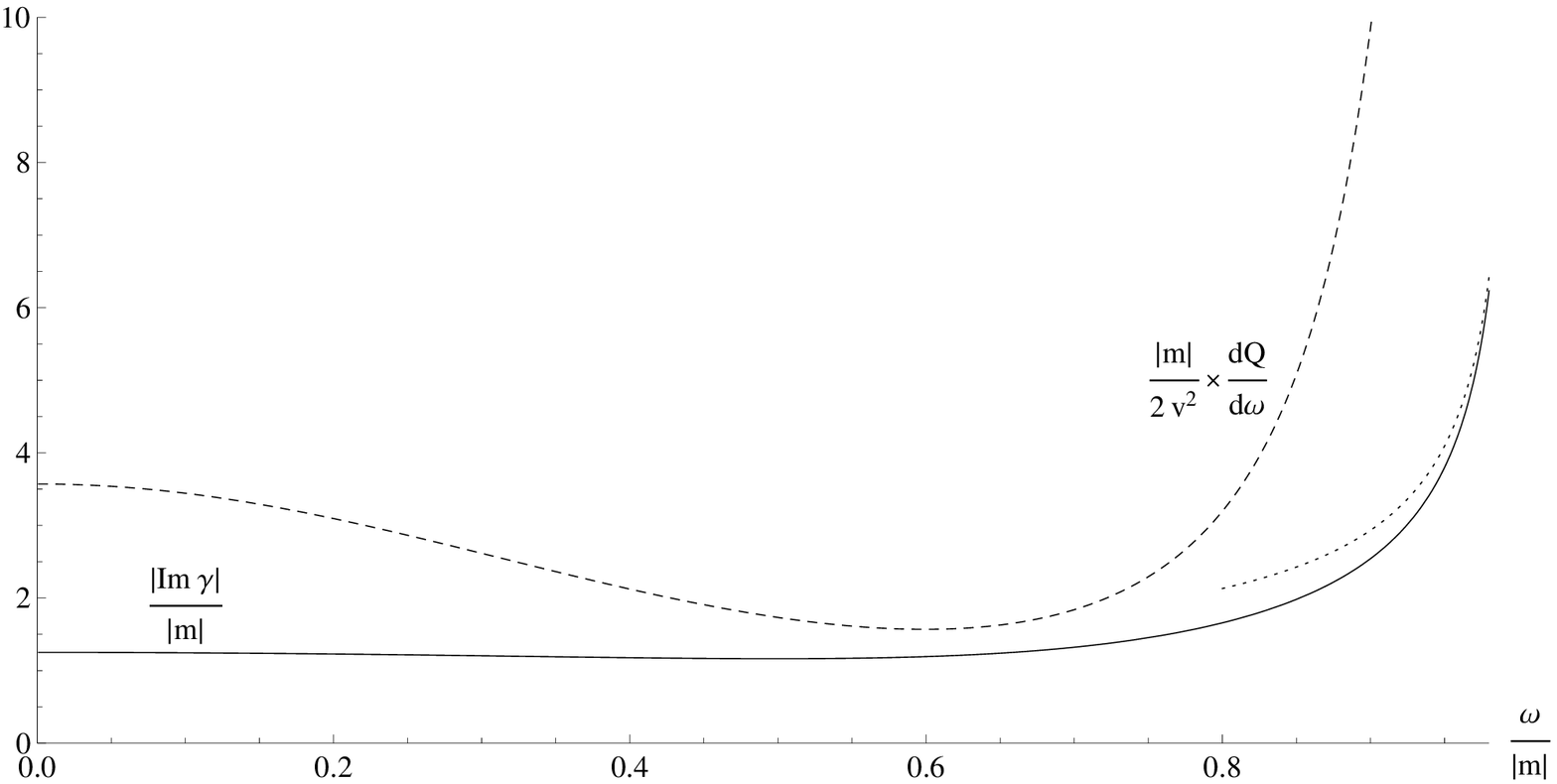}
\caption{The value of $|{\rm Im} \gamma|$ as a function of
$\omega$ for the case $m^2<0$, $M/|m|=1$. Solid line -- numerical
calculations; dotted line -- the function $\xi|F(\omega)|/|m|$.
The dashed line corresponds to the function $\frac{dQ}{d\omega}$.}
\label{gamma_Q2}
\end{figure}
The results of our analysis also show that Q-ball solutions with
minimal energy are always classically stable.

The numerical search for exponentially growing modes was also made
for the case $m^2<0$. We restricted ourselves only to examining
the even modes with $\textrm{Re}\gamma=0$. The results of the
numerical search for ${\rm Im}\gamma$ for even excitations
together with the function $\frac{dQ}{d\omega}$ are presented in
Fig.~\ref{gamma_Q1} and Fig.~\ref{gamma_Q2}. We see that in the
regions where $\frac{dQ}{d\omega}<0$ exponentially growing modes
are absent, whereas such modes exist if $\frac{dQ}{d\omega}>0$.
The latter means that Q-ball solutions in the phase presented on
the right plot in Fig.~\ref{E_Q1dim-2} are always classically
unstable, because $\frac{d^{2}E}{dQ^2}=\frac{d\omega}{dQ}>0$ in
this case. This result coincides with the results of numerical
analysis in the model of \cite{Anderson:1970et} in
$(3+1)$-dimensional space-time with an analogous $E(Q)$
dependence.

In the limit of large charges this instability can be obtained
analytically in all three cases ($m^2>0$, $m=0$, $m^2<0$). Let us
show it explicitly. Suppose that $\gamma$ is purely imaginary and
$|\gamma|\gg M$. In this case equation (\ref{perturb}) can be
rewritten as
\begin{equation}\label{perturb-unstab}
-|\gamma|^2\eta+\eta''-F\delta(|z|-R)
\left(\eta+\eta^{*}\right)\approx 0,
\end{equation}
which suggests that the function $\eta$ can be chosen to be real.
The even solution to equation (\ref{perturb-unstab}) takes the
form
\begin{eqnarray}
\eta(z)&=&\alpha e^{|\gamma| t}\cosh(|\gamma|z),\qquad |z|<R,\\
\eta(z)&=&\beta e^{|\gamma| t}e^{-|\gamma z|},\qquad |z|>R.
\end{eqnarray}
The continuity of $\eta$ at the points $z=\pm R$ and discontinuity
of its first derivative at these points (see equation
(\ref{matching})) results in
\begin{equation}\label{unst-gamma-analyt}
|\gamma|\left(\tanh(|\gamma|R)+1\right)=-2F.
\end{equation}
Let us parametrize $|\gamma|$ as
\begin{equation}\label{F}
|\gamma|=-\xi F
\end{equation}
For the solutions at large $Q$, which are supposed to be unstable,
we have $\omega\sim M$ and
\begin{equation}\label{R}
R\approx \frac{\sqrt{M^2-\omega^2}}{M^2-m^2}.
\end{equation}
Substituting (\ref{F}) and (\ref{R}) into
(\ref{unst-gamma-analyt}) we arrive at
\begin{equation}
\xi\left(\tanh(\xi/2)+1\right)=2,
\end{equation}
which results in $\xi\approx 1.2785$. The function
$\xi|F(\omega)|/|m|$ together with the numerical solution for
$|{\rm Im} \gamma|$ is presented in Fig.~\ref{gamma_Q},
Fig.~\ref{gamma_Q1} and Fig.~\ref{gamma_Q2}.

An important remark concerning the validity of the linear
approximation is in order. We examine classical perturbations, and
if the scalar field potential is smooth, we can consider linear
approximation without any restrictions (we can always choose small
enough amplitude of perturbations which does not destroy linear
approximation). Our case is rather nonstandard because of the
Heaviside step function in the scalar field potential. The
$\theta$ function can be easily regularized as
\begin{equation}
\theta\left(\frac{\phi^*\phi}{v^2}-1\right)\to\frac{1}{2}\left(1+\tanh{\left[
\alpha\left(\frac{\phi^*\phi}{v^2}-1\right)\right]}\right),
\label{regularization}
\end{equation}
where $\alpha$ is a large dimensionless parameter. In the linear
approximation it can be expanded as
\begin{equation}
\frac{1}{2}\left(1+\tanh{\left[
\alpha\left(\frac{\phi^*\phi}{v^2}-1\right)\right]}\right)\approx
\frac{1}{2}\left(1+\tanh{\left[
\alpha\left(\frac{f^{2}}{v^2}-1\right)\right]}\right)+\frac{\alpha
f(\eta+\eta^{*})}{2v^{2}\cosh^{2}\left(\alpha\left(\frac{f^{2}}{v^2}-1\right)\right)}.
\end{equation}
Let us take perturbations at the points $|z|=R$. It is clear that
the relation $\frac{\alpha(\eta+\eta^{*})}{v}\ll 1$ should hold in
order not to break down the linear approximation. In the limit
$\alpha\to\infty$ (which is exactly the case for which the
linearized equation (\ref{perturb}) was obtained), this relation
obviously leads to the constraint $\eta+\eta^{*}|_{|z|=R}=0$,
otherwise formally we get the breakdown of the linear
approximation. Thus, let us consider the linearized theory with
the constraint
\begin{equation}\label{constraint}
\eta+\eta^{*}|_{|z|=R}=0.
\end{equation}
Substituting (\ref{subst-lin}) into (\ref{constraint}) and taking
into account the fact that the relation (\ref{constraint}) should
hold at any moment of time, we get
\begin{equation}\label{eq48}
\psi_{2}(\pm R)=-\psi_{1}(\pm R).
\end{equation}
Now we take the first equation of (\ref{stab_system}). Taking into
account (\ref{eq48}), it can be rewritten as
\begin{equation}\label{L1}
\hat L\psi_{1}=(\omega+\gamma)^2\psi_{1}
\end{equation}
and
\begin{equation}\label{L2}
\hat L\psi_{1}^{*}=(\omega+\gamma^{*})^2\psi_{1}^{*},
\end{equation}
where the operator $\hat L$ is defined as $\hat
L=-\partial^2_z+U(z)$. Multiplying (\ref{L1}) by $\psi_{1}^{*}$,
integrating over the coordinate $z$ and using (\ref{L2}), we get
\begin{equation}
(\omega+\gamma)^2=(\omega+\gamma^{*})^2.
\end{equation}
The latter equation has two solutions. The first one leads to
$\textrm{Im}\gamma=0$, which means that exponentially growing
modes, indicating instability, are absent in this case. The second
solution is $\gamma=-\omega+i\gamma_{i}$, where
$\gamma_{i}=\textrm{Im}\gamma$. Using (\ref{L1}), we easily obtain
\begin{equation}
\langle\psi_{1}|\hat
L|\psi_{1}\rangle=-\gamma_{i}^2\langle\psi_{1}|\psi_{1}\rangle.
\end{equation}
But this relation can not be fulfilled, because
$\langle\psi_{1}|\hat L|\psi_{1}\rangle$ is nonnegative for any
$\psi_{1}$. Indeed, the eigenfunction of the operator $\hat L$,
corresponding to the eigenvalue $\omega^{2}>0$, is $f(z)$. This
eigenfunction has no nodes, which means that it is the lowest mode
and thus all other eigenvalues are also larger than zero, leading
to non-negativity of $\langle\psi_{1}|\hat L|\psi_{1}\rangle$.

Analogous calculations, leading to the same result, can be
performed for the function $\psi_2$. Thus, we have shown that
formally there are no modes indicating instability if constraint
(\ref{constraint}) is imposed. Meanwhile, this constraint is a
consequence of the existence of a generalized function in our
potential. A physically reasonable potential should be smooth,
leading to nonzero, though sometimes very small, amplitude of
perturbations. It this case in the linearized equations of motion
(\ref{perturb}) delta-function should be replaced by some smooth
function containing some parameter of regularization $\alpha$
(like the one in (\ref{regularization})). For very large, but
finite $\alpha$, the linearized equation of motion, as well as
corresponding solutions for perturbations and eigenvalues
$\gamma$, may look very similar to those in the $\alpha\to\infty$
limiting case. Of course, most probably such an equation can not
be solved analytically. But the difference between the exact
solution and the solution in the $\alpha\to\infty$ limiting case
is controlled by the parameter $\sim 1/\alpha$, and for very large
$\alpha$ the corrections to the case $\alpha\to\infty$ are
supposed to be very small. This reasoning justifies the use of the
"relaxed" linearized theory described by equation (\ref{perturb}),
without constraint (\ref{constraint}). This "relaxed" theory
allows one to see what could happen with the stability in a more
realistic case of a smooth scalar field potential.

\section{Conclusion}
In the present paper we discussed Q-ball solutions in a model
describing complex scalar field with a piecewise parabolic
potential (\ref{potential}). It was shown that due to the
simplicity of the potential Q-ball solutions can be found
analytically, which really simplifies the analysis of their
properties -- the spectra of solutions were obtained analytically
in (3+1)- and (1+1)-dimensional space-times for different values
of the parameters of the model. For $\omega\to M$ the Q-ball
solutions are very close to the condensate line $E=MQ$ and this
fact can be interesting for examining the Q-ball production. It
should be noted that the Q-ball solutions presented in this paper
can not be described by the standard thin-wall approximation.

The stability of the obtained solutions was also examined. In the
simpler theory in $(1+1)$-dimensional space-time it was shown
explicitly that solutions with $\frac{dQ}{d\omega}>0$ are always
unstable, whereas solutions $\frac{dQ}{d\omega}<0$ do not contain
exponentially growing modes, at least of the simplest form $\sim
e^{|\gamma|t}$.

Though the potential (\ref{potential}) is very simple, it provides
a variety of Q-ball solutions of different types. We hope that the
existence of exact and simple analytical Q-ball solutions with
known properties (such as their stability and $E(Q)$ dependence)
allows one to consider the model, presented in this paper, as a
useful tool for examining different phenomenological scenarios
involving Q-balls.

\section*{Acknowledgements}
The authors are grateful to S. Demidov, D. Gorbunov, M. Libanov,
V. Rubakov, S. Sibiryakov and I. Volobuev for discussions. The
work was supported by grant of Russian Ministry of Education and
Science (agreement No. 8412). The work of E.Y.N. was supported in
part by grant NS-5590.2012.2 of the President of Russian
Federation and by RFBR grant 13-02-01127a. The work of M.N.S. was
supported in part by grant NS-3920.2012.2 of the President of
Russian Federation and by RFBR grant 12-02-93108-CNRSL-a.

\section{Appendix A}
Let us prove that inequality $E>\omega Q$ holds for any Q-ball
solution. Consider (D+1)-dimensional space-time. According to
(\ref{solution-form}) we can use the time-independent effective
action
\begin{equation}
S_{eff}=\int
d^{D}x\left(\omega^{2}f^{2}-\partial_{i}f\partial_{i}f-V(f)\right),
\end{equation}
where $i=1,...,D$, instead of the initial one. Suppose that there
exists a solution $f(\vec x)$ to the corresponding equation of
motion. Let us apply the scale transformation $f(\vec x)\to
f_{\lambda}(\vec x)=f(\lambda\vec x)$ to this solution and
substitute $f_{\lambda}(\vec x)$ into the effective action instead
of $f(\vec x)$ (this technique was used in \cite{Derrick} to show
the absence of time-independent soliton solutions in some models
with a nonlinear scalar field). Now it takes the form
\begin{eqnarray}
S_{eff}^{\lambda}=\int d^{D}x\left(\omega^{2}f^{2}(\lambda\vec
x)-\lambda^{2}\frac{\partial}{\partial(\lambda
x^{i})}f(\lambda\vec x)\frac{\partial}{\partial(\lambda
x^{i})}f(\lambda\vec x)-V(f(\lambda\vec x))\right)\\
\nonumber=\int
\frac{1}{\lambda^{D}}d^{D}x\left(\omega^{2}f^{2}-\lambda^{2}\partial_{i}f\partial_{i}f-V(f)\right),
\end{eqnarray}
where we have passed to the new variables $\lambda \vec x\to\vec
x$ in the second integral. According to the principle of least
action we have
\begin{equation}
\frac{S_{eff}^{\lambda}}{d\lambda}\biggl|_{\lambda=1}=0
\end{equation}
and thus
\begin{equation}
\int d^{D}xV(f)=\omega^{2}\int d^{D}xf^{2}-\frac{D-2}{D}\int
d^{D}x\partial_{i}f\partial_{i}f.
\end{equation}
Substituting the latter equation into the definition of energy and
taking into account the definition of charge $Q=2\omega \int
d^{D}xf^{2}$, we get
\begin{equation}\label{EomegaQbigger}
E=\int
d^{D}x\left(\omega^{2}f^{2}+\partial_{i}f\partial_{i}f+V(f)\right)=\omega
Q+\frac{2}{D}\int d^{D}x\partial_{i}f\partial_{i}f>\omega Q.
\end{equation}

\section{Appendix B}
Let us show that if the condition $d^{2}E/dQ^{2}<0$ is fulfilled,
then the Q-ball is stable against fission, i.e. against a decay
into Q-balls with smaller charges. Suppose $E(Q)$, where $Q$ is
supposed to be nonnegative, is a positive monotonically increasing
function in the region of charges we are interested in. In this
case condition $d^{2}E/dQ^{2}<0$ means that
\begin{equation}\label{q-decay}
\frac{E(Q)}{dQ}\biggl|_{Q=\tilde
Q_{1}}>\frac{E(Q)}{dQ}\biggl|_{Q=\tilde Q_{1}+\tilde Q}
\end{equation}
for $\tilde Q>0$. Let us integrate this inequality over the
coordinate $\tilde Q_{1}$ in the region $[Q_{1},Q_{2}]$. We get
\begin{equation}
E(Q_{2})-E(Q_{1})> E(Q_{2}+\tilde Q)-E(Q_{1}+\tilde Q),
\end{equation}
which can be rewritten as
\begin{equation}\label{q-decay1}
E(Q_{2})+E(Q_{1}+\tilde Q)> E(Q_{2}+\tilde Q)+E(Q_{1}),
\end{equation}
If we take $Q_{1}=0$ and if $E(0)$=0, then (\ref{q-decay1}) leads
to
\begin{equation}\label{q-decay2}
E(Q_{2})+E(\tilde Q)> E(\tilde Q+Q_{2}),
\end{equation}
which means that Q-ball fission is energetically forbidden. But in
many models including the one discussed in the present paper there
exists a minimal charge $Q_{min}\ne 0$ such that
$E(Q_{min})=E_{min}\ne 0$. In such a case, one can try to redefine
the function $E(Q)$ in the region $[0,Q_{min}]$ in order to get an
auxiliary function $E_{aux}(Q)$: $E_{aux}(0)=0$, $E_{aux}(Q)$ is a
smooth monotonically increasing function for $Q>0$,
$d^{2}E_{aux}(Q)/dQ^{2}<0$ and $E_{aux}(Q)=E(Q)$ for $Q\ge
Q_{min}$. If it is possible to construct the function
$E_{aux}(Q)$, then inequality (\ref{q-decay2}) is valid for
$\tilde Q, Q_{2}\ge 0$ and, consequently, for $\tilde Q, Q_{2}\ge
Q_{min}$ (of course, any Q-ball with the charge $Q<2Q_{min}$ is
always stable against fission regardless of the sign of
$d^{2}E/dQ^{2}$).

\begin{figure}[h!]
\includegraphics[width=6.5in]{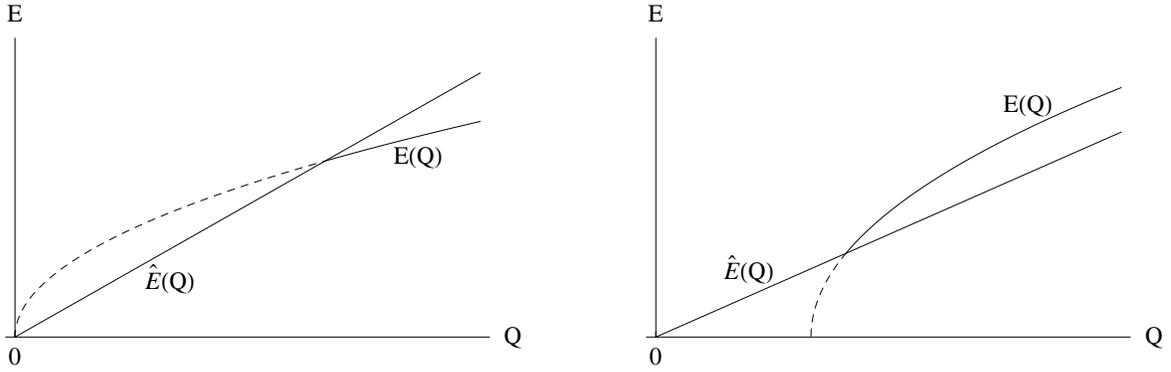}
\caption{Solid lines correspond to functions $E(Q)$ and $\hat
E(Q)$, dashed line corresponds to continuation of the function
$E(Q)$ (i.e., to the function $E_{aux}(Q)$).} \label{extrafig}
\end{figure}

To show that one can always construct such an auxiliary function
for a Q-ball let us consider the function $\hat E(Q)=\tilde\omega
Q$, where the constant $\tilde\omega$ is defined by
$\tilde\omega=E_{min}/Q_{min}$. Recall that
$\frac{dE}{dQ}\bigl|_{Q=Q_{min}}=\omega_{min}$. It is evident that
if $\frac{d\hat
E}{dQ}\bigl|_{Q=Q_{min}}=\tilde\omega>\omega_{min}=\frac{dE}{dQ}\bigl|_{Q=Q_{min}}$
then one can always construct a function $E_{aux}(Q)$, otherwise
it is impossible, see examples in Fig.~\ref{extrafig}.

Using equation (\ref{EomegaQbigger}) we get
\begin{equation}
E_{min}=\omega_{min}Q_{min}+\frac{2}{D}\int
d^{D}x\partial_{i}f\partial_{i}f=\tilde\omega Q_{min}
\end{equation}
and thus $\tilde\omega>\omega_{min}$. The latter means that we can
always construct a function $E_{aux}(Q)$ and inequality
(\ref{q-decay2}) fulfills if $d^{2}E/dQ^{2}< 0$ is valid.

\end{document}